\documentclass[acmtog]{acmart}

\usepackage{algorithm2e}
\usepackage[noend]{algpseudocode}
\usepackage{amsmath}

\usepackage{booktabs}
\usepackage{siunitx}
\usepackage{listings}
\usepackage{enumitem}
\usepackage{flushend}
\usepackage{subfig}
\usepackage{caption}
\newcommand{\xvar}[1]{\textsf{#1}}

\newcommand{\xvbox}[2]{\makebox[#1][l]{#2}}

\SetAlFnt{\small} %\footnotesize

\SetAlCapSty{xAlCapSty}

\SetCommentSty{xCommentSty}

\SetNlSty{mynlfont}{}{} %\scriptsize

\LinesNumbered

\SetSideCommentRight

\DontPrintSemicolon

\RestyleAlgo{algoruled}

\newcommand\setrow[1]{\gdef\rowmac{#1}#1\ignorespaces}
\newcommand\clearrow{\global\let\rowmac\relax}
\clearrow

\AtBeginDocument{%
  \providecommand\BibTeX{{%
    \normalfont B\kern-0.5em{\scshape i\kern-0.25em b}\kern-0.8em\TeX}}}

\setcopyright{acmcopyright}
\copyrightyear{2018}
\acmYear{2018}
\acmDOI{XXXXXXX.XXXXXXX}

\acmJournal{TOG}
\acmVolume{37}
\acmNumber{4}
\acmArticle{111}
\acmMonth{8}

\begin{document}

\title{Blockchain-based and Fuzzy Logic-enabled False Data Discovery for the Intelligent Autonomous Vehicular System}


\author{Ziaur Rahman}
\authornote{Corresponding Author}
\email{zia@iut-dhaka.edu}
\orcid{0000-0002-7759-34}
\author{Xun Yi}
\email{xun.yi@rmit.edu.au}
\author{Ibrahim Khalil}
% \authornotemark[2]
\email{ibrahim.khalil@rmit.edu.au}
\affiliation{%
  \institution{RMIT University}
  \streetaddress{414-418 Swnston St.}
  \city{Melbourne}
  \state{VIC}
  \country{Australia}
  \postcode{3000}
}

\author{Adnan Anwar}
\email{adnan.anwar@deakin.edu.au}
\author{Shantanu Pal}
\email{shantanu.pal@deakin.edu.au}
\affiliation{%
  \institution{Deakin University}
  \streetaddress{221 Burwood Hwy}
  \city{Melbourne}
  \state{VIC}
  \country{Australia}
  \postcode{3000}
  }
  
\renewcommand{\shortauthors}{Rahman and Xun, et al.}

\begin{abstract}
  Since the beginning of this decade, several incidents report that false data injection attacks targeting intelligent connected vehicles cause huge industrial damage and loss of lives. Data Theft, Flooding, Fuzzing, Hijacking, Malware Spoofing and Advanced Persistent Threats have been immensely growing attack that leads to end-user conflict by abolishing trust on autonomous vehicle. Looking after those sensitive data that contributes to measure the localisation factors of the vehicle, conventional centralised techniques can be misused to update the legitimate vehicular status maliciously. As investigated, the existing centralized false data detection approach based on state and likelihood estimation has a reprehensible trade-off in terms of accuracy, trust, cost, and efficiency. Blockchain with Fuzzy-logic Intelligence has shown its potential to solve localisation issues, trust and false data detection challenges encountered by today's autonomous vehicular system. The proposed  Blockchain-based fuzzy solution demonstrates a novel false data detection and reputation preservation technique. The illustrated proposed model filters false and anomalous data based on the vehicles' rules and behaviours. Besides improving the detection accuracy and eliminating the single point of failure, the contributions include appropriating fuzzy AI functions within the Road-side Unit node before authorizing status data by a Blockchain network. Finally, thorough experimental evaluation validates the effectiveness of the proposed model. 
\end{abstract}




\received{27 March 2023}
\received[revised]{April 2023}
\received[accepted]{June 2023}

\maketitle

\section{Introduction}
The world has experienced an appealing technological rise of intelligence-connected vehicular cyber-physical systems (CPS). According to the Internet Crime Complaint Center (IC3) of the United States(US) Federal Bureau of Investigation (FBI), 95\% of the recorded breaches targeted critical  infrastructure, such as sensor-enabled CPS. In July 2020, the Texas state power grid system was hacked, and the attacker tried to spoof the system’s monitoring tools to inject false data to bully the whole system. This was not the first time; a similar attack occurred in December 2015, when an attack on Ukraine’s power grid caused a massive blackout. These recent incidents reveal that the CPSs, including intelligence vehicles, are extremely vulnerable to False Data Injection (FDI). Blockchain has immense potential to secure the Vehicular CPS to protect it from injecting inaccurate data from neighbouring vehicles. Fuzzy logic and rule-based techniques can potentially discover data anomalies based on system behaviours. Instead of centralised monitoring, distributed and transparent control by the vehicle owners and roads and highways authorities, automatic false data detection can be advantageous if reliable techniques are involved. 

\begin{figure}[ht]
  \centering
  \includegraphics[width=\linewidth]{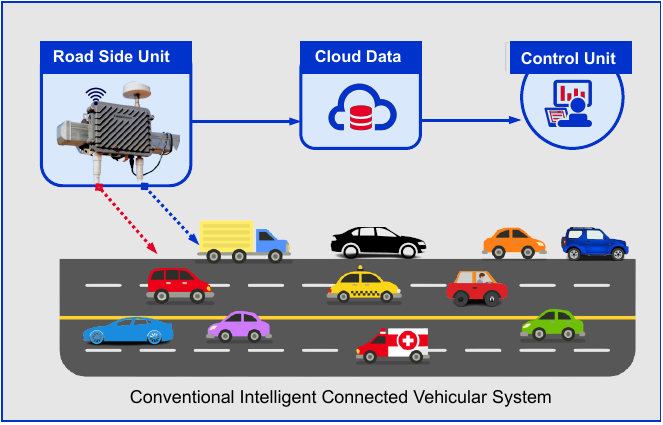}
  \caption{Conventional Vehicular Data Communication Flaw. The vehicles communicate with the nearby Roadside unit for an update. RUSU is connected with a Cloud-driven data centre controlled by the Roads-highway authority.}
  \label{fig1}
\end{figure}

An FDI attack is an unprecedented attack that often raises conflict against the reliable operation of the Vehicular CPS. Adding false data or misguiding an autonomous vehicle measurement may occur for different reasons. Any unauthorised intermediaries or even trusted neighbouring vehicles, intentionally or mistakenly, can inject malicious data. If the system control is maintained based on the trust employed through a TTP (i.e., service provider), the potential threat rises exponentially \cite{i4}. Accordingly, preserving incorrect or vehicular measurement data can be utterly misleading. Conventionally, the data-associated with the IAVS is maintained by a cloud from the provider side. Autonomous Vehicular stakeholders can see their contributions but barely have any control authority. At this point, if any vehicle is compromised with misleading data, the sole responsibility is on those entitled to control the system. In addition to data forgery, negligible and erroneous data may appear due to technical errors that deserve proper preservation for extensive record keeping and monitoring. In the IAVS, this status history often constructs a reputation that is necessarily important for further decision-making, cost measuring and further localisation and measurement. Figure \ref{fig1} shows the conventional network infrastructure of vehicular Cyber-physical systems. 

% \begin{figure}[ht]
% \centering
% \includegraphics[width=4.7 in]{Figure/pf111.pdf}
% \caption{A) False Data Injection (FDI) in a Supervisory Control and Data Acquisition (SCADA)–controlled Smart Grid CPS, B) False Data–Detection (FDD) Challenges.}
% \label{Ch3_Fig1}
% \end{figure}

\subsection{Challenges and Perspectives}
As the existing detection approaches demand proper revision to ensure transparency and accuracy, the research community has expressed deep concern for convincing solutions. However, blockchain has proven its ability to preserve transparent data transmission and sharing generated from a distributed network with the desired anonymity and immutability. They work through adaptable consensus and smart contract mechanisms \cite{cons}. A false data attack is a kind of attack targeting the autonomous vehicle that causes different disruptions including localisation and further estimation issues. In false data attacks, misleading information is allegedly appended to one of the major operational modules \cite{energy}. Therefore, detecting infected data and assessing how much data of IAVS is compromised should be done confidentially and securely.

Due to its immutable, efficient, reliable and enormously accessible behaviours, blockchain can be an exciting solution to this FDI and transparency problem \cite{li2018blockchain}. This article proposed a false data discovery and preservation technique. Beised, a reputation-building process is proposedly integrated. To secure status data while travelling from sensors to the blockchain ledger, the design incorporated a customised digital signature mechanism, fuzzy rule–based detection accuracy that works by following an infected data detection algorithm \cite{fuz1} \cite{fuz2}. Further, another functional algorithm was designed to communicate with the blockchain ledger. The fuzzy-based detection methods show convincing accuracy, and the blockchain-aligned reputation preservation process brings a transparent and secure outlook for IAVS management \cite{li2017distributed}.

\subsection{Contributions and Organizations}
This work was motivated to address and demonstrate a blockchain and AI-enabled false data–detection and reputation preservation for the sensor-enabled, intelligent Cyber-physical system especially targeting IAVS. The specific contributions of this work are as follows.

\begin{itemize}
\item The proposed Fuzzy logic-enabled false data–discovery technique can filter data anomalies based on the behaviour rules. As described in the respective sections, the fuzzy-based model has higher effectiveness in terms of cost and security. 
\item The proposed model incorporates a novel reputation preservation mechanism based on infected vehicles that potentially generate false or misleading data. The reputation status of the measurement units helps other autonomous vehicles to be aware of the devices and protect the system from being misled. 
\item Blockchain-based transaction verification ensures collaboratively built trust and security rather than relying on a single party. It eliminates PKI-driven cloud and centralised systems to protect the IAVS. 
\end{itemize}

% Therefore, the rest of the article is organised as follows. Blockchain and AI-aligned approaches to false data detection and preservation and the security assumption and trust model. Section 3.3 explains fuzzy rule specification based on the smart grid’s stable behaviour and addresses a novel reputation-updating algorithm. Section 3.4 illustrates design concepts and the system model. Section 3.5 discusses FDD and reputation update mechanism and Section 3.6 discusses the device registration and communication process, including the respective algorithms. Section 3.7 uses graphical representations to justify the claimed contributions.

\section{Background and Related Works}
In this section relevant background knowledge on Blockchain technology, Fuzzy AI technique, autonomous vehicular technology and Related works are presented. Table \ref{notations} depicts the technical terms, notations and respective abbreviations frequently used throughout the paper.

\begin{table}
\centering
\begin{tabular}{ll}
\toprule 
\textbf{Terms} & \textbf{Elaboration \& Description} \\  \midrule
ADAS & Advanced Driver Assistance Systems \\
AI & Artificial Intelligence \\
BC & Blockchain \\
BFT & Byzantine Fault Tolerance \\
BTC & Bitcoin \\
CC & Chaincode \\
CFT & Crash Fault Tolerance \\
CPS &  Cyber-physical System \\
ETH & Ethereum \\
FDD & False Data Detection \\
FL & Fuzzy Logic \\
HLF & Hyperledger Fabric \\
IAVS & Intelligence Autonomous Vehicular Systems \\
LiDAR & Light Detection and Ranging \\
MF & Membership Functions \\
MSP & Membership Service Provider \\
ML & Machine Learning \\
P2P & Peer to Peer Network \\
RADAR & Radio Detection And Ranging \\
RSU & Road Side Unit \\
SC & Smart Contract \\  
SPOF & Single-point of Failure \\
\bottomrule
\end{tabular}
\caption{\label{tab:widgets2}Technical Terminology along with its notation entries and abbreviation in alphabetic order}
\label{notations}
\end{table}

\subsection{Blockchain Technology for the Vehicular CPS}
\begin{figure}[ht!]
\centering
\includegraphics[width=\linewidth]{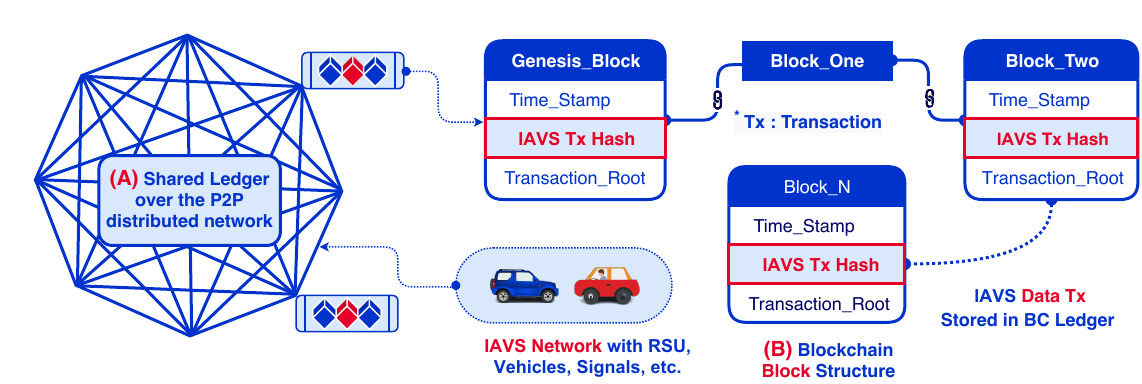}
\caption{ Sample Blockchain Structure Consisting of IAVS Transactions ($Tx$). A) Peer-to-peer (P2P) network of vehicular CPS where blockchain peers communicate and B) Blockchain block structure}
\label{Ch3_Fig2}
\end{figure}

The emerging blockchain technology has immense potential to secure and enhance autonomous driving operations and management. Because of its self-governing smart contract protocols and consensus-driven block verification, its integration into the IAVS increases data and communication integrity and security \cite{surbc2}. As shown in Figure \ref{Ch3_Fig2}, blockchain is an expanding and unchangeable list of records consisting of connected blocks using a secure and immutable hash algorithm. The network works on the distributed P2P network constituted by IAVS components, such as moving vehicular sensors, LiDAR sensors etc. Unlike centralised cloud-driven services, blockchain ensures multiparty authorisation, which essentially eliminates SPOF \cite{survey3}. 

 Before storing a IAVS transaction in an associated ledger, it must be consented to by the contributory peers through a special process called the consensus mechanism. Earlier generation blockchain, such as Bitcoin and Ethereum, incorporated the PoW type of consensus, which is often criticised because of its significantly slower transaction processing rate. Based on the joining rights, blockchain can be either public or consortium, where only authorised users are allowed to join and contribute. Apart from PoW, CBC like Corda, HLF and Ripple incorporate fault tolerance consensus techniques (e.g., BFT and crash fault tolerance [CFT]). Crash fault tolerance excludes longer ID and transaction verification and, thus, has higher throughput and negligible DL \cite{li2018blockchain}. For example, for Bitcoin and Ethereum, the transaction processing rate (known throughput, transactions per second) ranges from 4 to 15 transactions per second, whereas HLF can process 3,000–20,000 transactions per second. By excluding computation-intensive validation, it eliminates conventional rewards or incentives, which makes CBC a great alternative for real-time and critical infrastructure, such as smart grids and IAVS \cite{gdpr, zia9, zia7}.

\subsection{FL for False Data Discovery}
Fuzzy logic is a form of AI reasoning that makes decisions in the same way as humans. Its computer-digestible logic block takes precise input and produces a definite output equivalent to real-world reasoning. IAVS follow particular rules and behaviour that can be logically translated into an input membership function (MF) of AI FL \cite{fuz1}. Thus, several MFs build intelligence together for a decision required for a particular IAVS. Unlike Boolean logic or probability theory, its decision-making process relies on the degrees of truth factor between $true$ and $false$. Although FL is based on the levels of probabilities of input variables towards the purposeful output, it is a subset of AI that can be trained using software, hardware or both. The fundamental FL architecture contains at least four components, including rule specification and MFs. Where an MF for a fuzzy set $f$ on the universe of discourse $y$ is defined as $\mu_f : y \: \rightarrow [ \, 0, 1 ] \,$. The advantages of FL system are as follows \cite{fuz2}.

\begin{itemize}
\item Mathematical concepts for FL reasoning are simple to implement and can be modified easily by revising the integrated rules.
\item Fuzzy logic systems can work dynamically with imprecise, anomalous input data. As a result, reasoning and decision-making can be made with fewer power constraints, reducing system deployment costs.
\end{itemize}

\begin{figure}[ht]
\centering
\includegraphics[width=\linewidth]{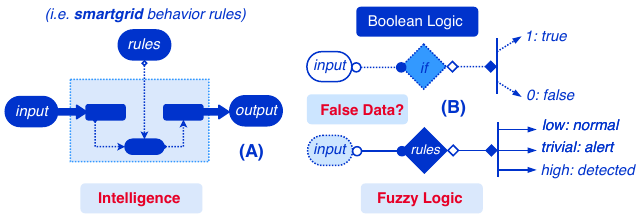}
\caption{Fuzzy logic (FL) components and salient characteristics that makes it distinct from its counterpart name boolean logic.}
\label{Ch3_Fig3}
\end{figure}

\vspace{\baselineskip}
\noindent Figure \ref{Ch3_Fig3} shows the basic components of FL and how it varies from Boolean logic. The rule can relate to any conditions or behaviours. For example, False data injection attack on a neighboring vehicle is doscovered when the Error ($E$) is larger than a threshold ($E_{threshold}$) and the Weight ($W$) for that source is lower than a threshold ($W_{threshold}$). The edge node trained with these autonomous vehicular  behaviours can detect data status and identify the source device if such conditions (as determined by MF) are not met. The considered IAVS rules will be explained in the forthcoming Section.

\subsection{Related Works} 
False Data detection in the cyber-physical system has attracted research community for a couple of years and a good number of works highlighted the importance of the stealthy FDI attacks~\cite{zia1, zia2}. \cite{zia3, zia4} proposed a secure model for data attack detection. The work above seems to have better performance as claimed through their simulation-based evaluation. In \cite{zia5} \cite{zia6, zia7} the authors proposed a monitoring system to determine the real-time occurrence of a disturbance in the voltage before suggesting a remedy in response. A group of researchers has recently made a private blockchain-based approach for local power consumption and generation without any trusted intermediaries. Another distributed ledger-driven effort based on smart-contract was explained by the authors to enhance the security and resilience of the energy CPS\cite{b4}. Apart from a distributed ledger, work done on distributing the host-based approach to detect FDI attacks by proposing novel False Data Detection (FDD) method, state estimation and performance reputation update with maximum likelihood algorithm\cite{zia1}. In their work, the authors have considered distributed host-based effort instead of the distributed ledger, and the rules assumed to evaluate seem to not exceed four host monitors. We have extracted sample from three different rules mostly on the autonomous vehicles's wegiht measurement, particle filtering, data fusion behaviours throughout our initial investigations, but even this number not seems to be portraying an entirely complex scenario of the IAVS. In our approach, we also have considered distributed network and instead of centralized monitoring. Distributed ledger both private and public blockchain have been incorporated. Another work done based on weak data attack arising due to stealthy and corrupted measurement seems to be done the experiments and demonstrated theoretical analysis before claiming their approach has less relative error \cite{b9}. A lightweight privacy-preserving technique for distributed RSU has also claimed the authentication speed \cite{b10}. 

\begin{table*}
\renewcommand{\arraystretch}{1.3}
\caption{Overview of recent research on IDSs for IoT applications}
\label{Table: Overview of recent research on IDSs for IoT applications}
\centering
\begin{tabular}{>{\rowmac}l>{\rowmac}c>{\rowmac}l>{\rowmac}l>{\rowmac}l>{\rowmac}c<{\clearrow}}
\toprule
\setrow{\bfseries}Referral Work & Crypto & Efficiency & Limitations & Application & Blockchain \\
\midrule
\cite{zia1} & $\surd$  & Low & IAVS Incompatible \ & Security &  $\times$ \\

\cite{zia2} & $\surd$  & Medium & Cost-intensive & Small-scale IoT &  $\surd$ \\

\cite{zia3} & $\surd$ & Low & Coalition attacks & Security &  $\times$ \\

\cite{zia4} & $\surd$  & Low & Chosen-text attack & Security &  $\times$ \\

\cite{zia4}& $\times$ & High & Cost-intensive & IoT &  $\times$ \\

\cite{IIoT} & $\times$  & Low & Com.-Intensive & Medical IoT &  $\times$ \\

\cite{gdpr} & $\times$  & Medium & High overhead & Vehicular Sensor &  $\times$ \\

%BC

\cite{Anwar201758} & $\surd$  & Medium & Not efficient & VANET &  $\surd$ \\

\cite{b2} & $\surd$  & High & Cost-intensive & Critical system &  $\surd$ \\

\cite{b3} & $\surd$  & Very High & Com.-intensive & Industrial IoT &  $\surd$ \\ \bottomrule

\end{tabular}
\end{table*}

\section{Design Concepts and System Model}
The proposed design concepts include three different components. Firstly, a fuzzy-based false detection technique in the Roadside Unit (RSU) end filters data before sending it to the blockchain network. Secondly, the blockchain authenticates data and the generating source devices through a certificateless and collaborative signing process \cite{multi, multisig}. Finally, only the verified data are stored in the storage node. The proposed framework was designed considering these salient features and incorporated a permissioned blockchain and DHT mechanism for demonstration and evaluation \cite{hlf, gdpr}. However, it conceptually supports the public type of blockchain and storage service. Figure \ref{Ch3_Fig4} shows the high-level view of the proposed detection and reputation preservation approach. The communication flow of the proposed system can be divided into three parts: discussed as follows.

\subsection{Vehcile to RSU Communication}
Autonomous vehicular CPS employs Sensor Fusion  such as Radars and LiDARs. Fusion sensors are connected to the global remote terminal unit via the V2R (Vehicle-to-RSU) network. Other IoT sensors are able to send data to the destination through the constituted edge devices, irrespective of whether sources are wired or wireless \cite{energy2, AnwarIoT}. Portion A of Figure \ref{Ch3_Fig4} shows the V2r communication where the proposed fuzzy rules work to detect data anomalies \cite{fuz1}.

\subsection{Blockhain Ensures Secure Data Transport}
Instead of cloud-driven systems, the edge data are authenticated via a blockchain network to reduce the chance of SPOF and centralised trust. The proposed solution incorporates a certificateless multisignature-based device and data authentication over the P2P network \cite{multisig, iot1}. The network can be either public or restricted; however, considering the high data processing time and DL, the proposed model constitutes CBC. It establishes secure communication with the IAVS RSU and MEMS and validates the transported data. Portion B of Figure \ref{Ch3_Fig4} depicts the secure data transport using blockchain. 

\subsection{Reputation Preservation and Storing}
\begin{figure} 
\centering
\includegraphics[width=\linewidth]{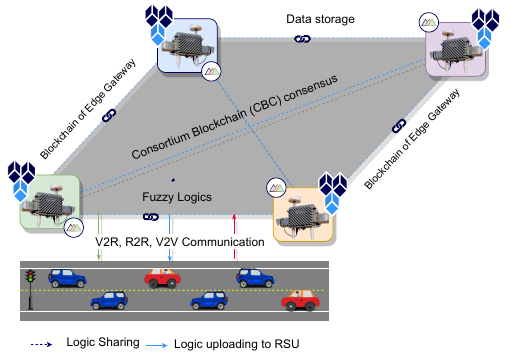}
\caption{High-level Representation of the Proposed Blockchain and Fuzzy Logic-aligned False Data–Detection and Reputation Preservation for the Intelligent Autonmous Vehicular System (IAVS). A) Source Vehicles send data through the edge gateway, B) the client vehicles submits the data transaction to the blockchain network (i.e., key generation and distribution [KGD] consortium) and C) upon successful verification, the data transaction and the detection status and reputation are recorded in the ledger, and data are stored in distributed hash table storage (e.g., interplanetary file system [IPFS] and Kademlia). }
\label{Ch3_Fig4}
\end{figure}

The reputation preservation algorithm works within the detection model to update the reputation of autonomous vehicles. Once the particular vehicle seem to be generating false data, it will update its individual status. The reputation and the data transaction are recorded in the blockchain ledger, and data are stored in off-chain storage. The proposed framework considered the salient features and incorporated DHTs, such as IPFS and Kademlia \cite{cons1}. Reputation preservation happens in the earlier portion; Portions A and C of Figure \ref{Ch3_Fig4} portrays the storage mechanism. However, storing data directly in the blockchain network, even in an encrypted form, threatens consumer or stakeholder privacy and does not comply with privacy standards \cite{zia8, zia9}. 

\vspace{\baselineskip}

\subsection{Threat Model}
The proposed model was designed based on the considered threat model. The decentralised blockchain ensures that an attacker cannot corrupt the consortium network. Any unauthorised peer or adversary cannot modify the blockchain ledgers, which implies that the resource is compromised. The threat model prevents impersonation by an unauthorised party or an adversary, as the associated multisignature cannot be tempered or forged. Therefore, security threats can be generalised into two broad categories. Firstly, an internal party or peer disguised in a Byzantine way has probably been granted access to IAVS data  \cite{cons3}. Secondly, an honest or trusted vehicle but its security credentials, such as private or decryption keys, are disclosed to an external adversary. Thus, the external party with the stolen access can bully the network. Blockchain smart contracts contain a token validation technique, which is refreshingly expired after a particular time or transaction, that protects the network from being compromised with the latest type of threat. However, the blockchain ledger will record the reputation of the malicious peers and block them temporarily or permanently. The BFT or CFT technique ensures the system runs smoothly, even after some peers have been suspended. Besides the security threats, the model considers the privacy of the IAVS and their data. The encryption and partial secret (PS) of the multisignature ensure pseudo-anonymity, whereas CBC only allows authorised peers, meeting the privacy challenges of the Autonomous Vehicular CPS \cite{cons}.  

\subsection{Trust Assumption}
The proposed model assumes that RSU constituting the blockchain network are honest or semi-honest. The model obviates the membership service provider (MSP) who has equivalent CA to PKI \cite{pki}. As investigated throughout the centralised cloud-driven approaches, it increases the chance of being compromised and SPOF. Besides, the elliptic curve–cryptographic primitives and hash function are assumed to be particularly secure. This means that attackers cannot extract keys using reverse exponentiation, break the hash algorithms or temper the multi-signature. In addition, the model considers that the data transfer occurs over an insecure network or internet. The next section discusses the proposed mechanism for false data detection and preservation  \cite{clKGC}.  

\section{False Data Discovery and Reputation Update}
As assumed that the IAVS usually operates on a normal stable status where the associated state parameters and variables differ in an interchangeably balanced manner. For example, the IAVS follows specific behaviours. Thus, any variable state changes due to a system fault cause corresponding state changes and produce anomalous data. However, data anomaly can be identified if variables change on one bus without affecting the parallel variables. 

In this paper, IAVS communicates with the RSU servers and publishes information about itself and its neighbours with a unique vehicle identity (ID). After the RSU server gathers the information from all the IAVS in a platoon, neighbouring vehicles information can be associated using the vehicle ID, and neighbouring unconnected vehicles information from multiple IAVS on-board sensors (i.e., Lidar or Radar) is assumed to be fused using a multi-source data association method so that each unconnected vehicle is also assigned with a specific vehicle ID. Therefore, by leveraging vehicle IDs, data for neighboring vehicles can be identified, and only neighbouring IAVS information will be used in the proposed solution.

\begin{figure}[ht]
\centering
\includegraphics[width= .7\linewidth]{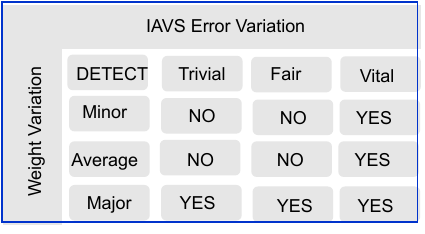}
\caption{Behavoural rule extraction and its corresponding fuzzy representation. A) Rule matrix for different status B) Rule specifications }
\label{Ch3_Fig5}
\end{figure}

% \RestyleAlgo{ruled}

% %% This is needed if you want to add comments in
% %% your algorithm with \Comment
% \SetKwComment{Comment}{/* }{ */}

% \begin{algorithm}[hbt!]
% \caption{An algorithm with caption}\label{alg:two}
% \KwData{$n \geq 0$}
% \KwResult{$y = x^n$}
% $y \gets 1$\;
% $X \gets x$\;
% $N \gets n$\;
% \While{$N \neq 0$}{
%   \eIf{$N$ is even}{
%     $X \gets X \times X$\;
%     $N \gets \frac{N}{2} $ \Comment*[r]{This is a comment}
%   }{\If{$N$ is odd}{
%       $y \gets y \times X$\;
%       $N \gets N - 1$\;
%     }
%   }
% }
% \end{algorithm}

\subsection{Rule Specifications}

When a IAVS is under usual operation, all of its state variables follow particular constraints and hold desired properties. 

% For example, power ($P$ ) meets the following conditions.

% \begin{itemize}
% \item $P_{min} < P^t < P_{max}$ power at time $t$ should very in an range of $(P_{main}, P_{max})$.
% \item $\mid P^t - P^{t-1}\mid \:< P_{\Delta} $ power variation at $t$ interval should be less than the threshold. 
% \end{itemize}

The following Table \ref{Ch3_Tab3} shows similar rules considered. These are some fundamental rule specifications to detect false data due to anomalous PMU activities.

% \begin{table}
%   \caption{Frequency of Special Characters}
%   \label{tab:freq}
%   \begin{tabular}{ccl}
%     \toprule
%     Non-English or Math&Frequency&Comments\\
%     \midrule
%     \O & 1 in 1,000& For Swedish names\\
%     $\pi$ & 1 in 5& Common in math\\
%     \$ & 4 in 5 & Used in business\\
%     $\Psi^2_1$ & 1 in 40,000& Unexplained usage\\
%   \bottomrule
% \end{tabular}
% \end{table}

% \begin{table*}
%   \caption{Some Typical Commands}
%   \label{tab:commands}
%   \begin{tabular}{ccl}
%     \toprule
%     Command &A Number & Comments\\
%     \midrule
%     \texttt{{\char'134}author} & 100& Author \\
%     \texttt{{\char'134}table}& 300 & For tables\\
%     \texttt{{\char'134}table*}& 400& For wider tables\\
%     \bottomrule
%   \end{tabular}
% \end{table*}

 \begin{table}
\centering
\caption{\label{Ch3_Tab3} Typical behaviour of IAVS rule examples}
\begin{tabular}{lll}
\toprule
Sl & Behaviour Rules & Variable Description \\\midrule
1 & $\Delta I_{k} > I_{Threshold}$ & Malicious data injected by a vehicle\\ 
2 & $P_E = P_F + P_M$ & $P_E$ from probability of ($P_M$) \& ($P_F$) \\ 
3 & $E_t > E_{Threshold}$ & Error larger than threshold \\ 
4 & $W_t < W_{Threshold}$ & Weight lesser than threshold \\ 
\bottomrule
\end{tabular}
\label{tab2}
\end{table}

The fuzzy rule specifications as explained in the next subsection considers following basic rules. Behavioural rules can be similarly specified for all other rules listed in the above Table. 

\begin{itemize}
\item a false data injection attack on a neighbouring vehicle is identified when the error ($E_t$) at time $t$ is larger than a threshold $E_{thershold}$
\item Weight at time $t$ for that source is lower than a threshold ($W_{Threshold}$)
\item $P_E = P_F + P_M$ means that Error obtained from probability of misdetection ($P_M$) \& the probability density function ($P_F$)
\end{itemize}

\vspace{\baselineskip}
\noindent The following Figure \ref{Ch3_Fig5} shows the rule-matrix that works to filter the data quality. First, it needs to classify the behaviour in different states \cite{b2}. For example, as per Rule 4 of Table \ref{tab2}, the variation of Error should not be always greater than a measured threshold. The threshold can be calculated following up the dynamic nature of the IAVS and previous records. However, as settled that the Weight variation at a time $t$ should be always less than the threshold. Considering the severity of the difference, it Fuzzy system classify, it as $minor$, $average$ or $major$. Similarly, for Error, it can be $trivial$, $fair$ or $vital$. Based on the rule matrix as demonstrated by Figure \ref{Ch3_Fig5}, the corresponding fuzzy rules are listed above. For example, if Error deviation is $trivial$ and Weight is $minor$ then the fuzzy system will not mark it as $NO$ and will send it to the blockchain peers for further processing. In different cases, it will either $YES$ data as anomalous or send it with a $warning$ flag \cite{b4, b5}. 

%Membership function (MF) for the variation of active power angle
% Membership function (MF) for the variation of voltage amplitude

\begin{figure}[ht]
    \centering
    \subfloat[\centering Test 1]{{\includegraphics[width=\linewidth]{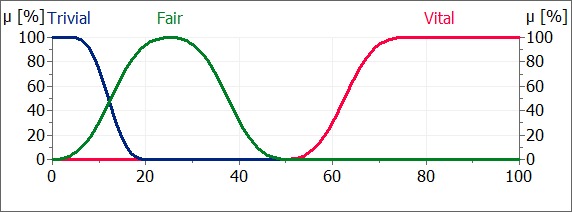} }}%
    \qquad
    \subfloat[\centering Test 2]{{\includegraphics[width=\linewidth]{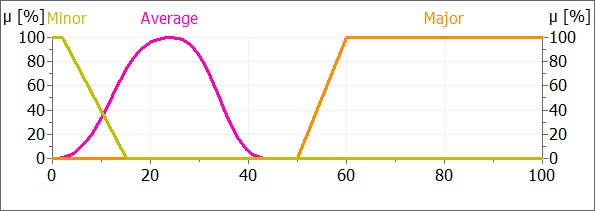} }}%
    \caption{Input MF definitions based on IAVS behavior rules}%
    \label{Ch3_Fig6}%
\end{figure}

\subsection{Defining Fuzzy Membership Function (MF)}
The graphical representation of fuzzy membership function (MF) shows how each point in the input space is mapped to the corresponding system status. FL modelling includes at least four components including rule specification and membership MFs. Where an MF for a fuzzy set $f$ on the universe of discourse $y$ is defined as $\mu_f : y \: \rightarrow [ \, 0, 1 ] \,$.  It quantifies the severity of MF element both in $x$ and $y$ axis where $x-$axis shows the universe of discourse and $y-axis$ represents the degree such as $trivial$, $minor$, $fair$ etc. within the variation range. As investigated, the accuracy varies as per type MF functions \cite{fuz1}. For example, if Error is implemented with a $triangular$ function, the detection varies from the $trapezoidal$ MF. Targeting the maximum throughput, the proposed evaluation runs with the $gaussain$ and its variant $SP-line$ MF. In a normalized $SP-line$ MF $\mu_i^m$ of order $m$ (degree $(m-1)$ for the fuzzy subset $[a,b]$ over $R$ (Real number range) the variation $\Delta:\: a = k_0 < k_1 <....<k_{n+1} = b$ as $\mu : [a,b]\rightarrow [0,1]$. Here $m_i$ is the multiplicity of the knot $k_i$.

\vspace{\baselineskip}
\noindent Figure \ref{Ch3_Fig6} depicts the respective membership of functions of based on the degree of variation of both Error and Weight angel as mentioned earlier. During the range selection of the demonstration, we changed ranges to a different level. For example, the following Figure \ref{Ch3_Fig6} shows that if the variation exceeds about 50\% then severity is classified as $vital$ for Error and $major$ for Wegiht deviation. However, based on the previous record of RSU the ranges could be varied to improve the system performance \cite{fuz2}. 

% \begin{figure}[ht]
% \centering
% \includegraphics[width= \linewidth]{Figure/Sub/pf7.png}
% \caption{Complete debugging scenario of the fuzzy detection system. It includes A)-C) Input and output membership functions(MF), D) fuzzy system model, E) debugging pane and F) extracted behaviour rules.}
% \label{Ch3_Fig7}
% \end{figure}

\vspace{\baselineskip}
\noindent Following a similar process, the output membership functions have been selected. The $gaussian$ and $SP-line$ seem to bring higher accuracy in comparison to $trapezoidal$ and $triangular$ MF. The threshold basically depends on the previous record of the Vehicular CPS, however, it has been finalized one-fourth ($25\%$) of the system's overall deviation. That means it verdicts the $detected$ if the average variation of Weight exceeds the Threshold. Sample false data detection after debugging the MF and its configured behaviour rules is discussed here. Here $A$ and $B$ are the input MF configuration based on $E_t$ as explained earlier \cite{fuz2}. $C$ depicts the corresponding output. For example, for a particular case, $W_t$ becomes varies within the Threshold then it detects the severity of the False Data is about $85\%$. In such a circumstance, the system as integrated in the edge gateway, will not allow sending the corresponding data transaction to further blockchain peers. Besides, it will update the reputation of source PMU and will include the latest status along with data transaction and source identities \cite{fuz1}. 

\subsection{RSU Reputation Updating}
The probability distribution function (PDF) can be applied to determine the system's reputation. For example, $\beta$ distribution seems to be promising for a collaborative detection model. Considering further smooth and secure preservation aligned with blockchain, the proposed model incorporates a novel reputation updating algorithm based on the degree of the detection level \cite{li2017distributed}.  

\subsubsection{Reputation Algorithm}
The \textbf{Alg}. \ref{Chap3_Alg1} takes input parameters from the previous detection phase. Parameters include system detection level and status which is either $true$ or $false$, identities of the PMU or any other similar sensors or MEMS along with the corresponding values of the membership variables,. The respective functions or subroutine initializes the required system parameters such as  initial reputation and associated values of the RSU.

\begin{algorithm}

% functions
\SetKwFunction{init}{init}
\SetKwFunction{setStatus}{setStatus}
\SetKwFunction{getStatus}{getStatus}
\SetKwFunction{updateRep}{updateRep}
\SetKwFunction{refresh}{refresh}

% input/ouput names
\SetKwInOut{Input}{Input}
\SetKwInOut{Output}{Output}

% caption
% \addto\captionsenglish{\renewcommand{\algorithmcfname{Alg.}}

\caption{IAVS RSU reputation updating on false data discovery.\label{Chap3_Alg1}}

\Input{%
		\xvbox{4mm}{$\xvar{S}$} -- status either $true$ or $false$ \\
		\xvbox{4mm}{$\xvar{R}$} -- reputation level  \\
		\xvbox{2mm}{$\xvar{L}$} -- RSU or sensor identities \\
		\xvbox{2mm}{$\xvar{A}$} -- Error \\
		\xvbox{2mm}{$\xvar{V}$} -- Weight \\
		\xvbox{2mm}{$\xvar{D}$} -- detection level 
         \tcc*{received data from fuzzy system}
	  }
\Output{%
		\xvbox{17mm}{$\xvar{($ID$,$R_t$, $D$)}$} -- returns after algorithm execution 
		} 

  \BlankLine % blank line for spacing
 
  \xvbox{2mm}{$\xvar{\init \; :=  (ID, \;R, \; D, \;$A$, \;$V$)}$}  \tcc*{initializes after fuzzy detection}
  
  \For{$\xvar{ID}  \leftarrow$ $\xvar{ID}_i$ \tcc*{for all identities $n\times ID_i$} }{
  
  \xvbox{2mm}{$\xvar{$R \leftarrow$ \getStatus(ID},D,R)$}  \tcc*{get requisite values ($ID$)}

  \If{$\xvar{(status}\; == \;true)$ }{

    \xvbox{2mm}{$\xvar{S $\leftarrow$ \updateRep(R, ID):}$}  \tcc*{update PMU or sensor reputation}
  }
  }
\end{algorithm}

\begin{figure}[ht]
\centering
\includegraphics[width= \linewidth]{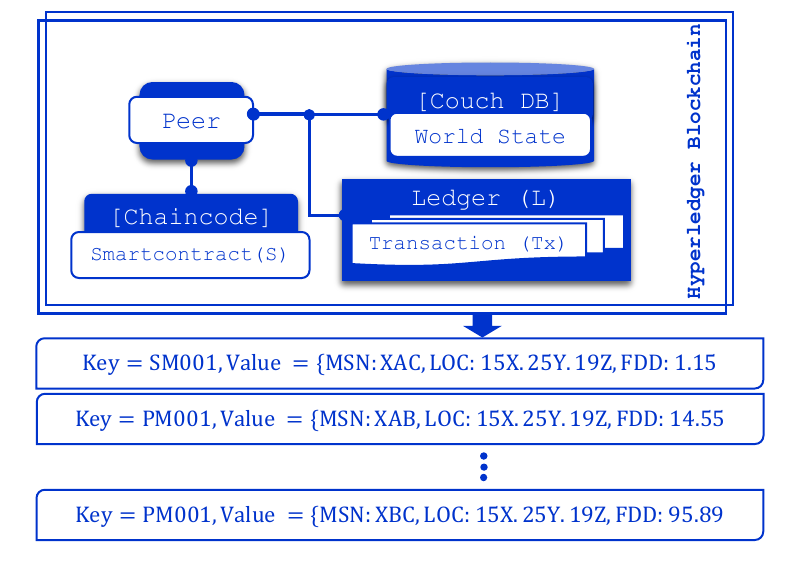}
\caption{Sample IAVS transaction belongs to the ledger and typical world state database (i.e. Couch DB) with interaction with chaincode smart contract.   }
\label{Ch3_Fig8}
\end{figure}

\vspace{\baselineskip}
\noindent Once values are set, the algorithm checks if and only if the status is true or false. It updates the particular RSU (identified with the ID) status and exit process if any false data are found within the system, as predicted by the earlier detection phase. After finalising the detection and reputation updating process, the data are ready to be sent to the blockchain for further verification and storage. The next section discusses how the IAVS detection status and reputation level are validated by the associated blockchain network and successfully stored for future maintenance and preservation. Next, the IAVS data and the corresponding reputation need to be transformed into a blockchain transaction. Figure \ref{Ch3_Fig8} shows the sample IAVS transaction to be transported over the internet.

\section{Blockchain Verification and Data Preservation}
In the proposed IAVS data verification and preservation process, the Consortium blockchain plays an indispensable role. The PMU needs to get registered with the blockchain-based key generating and distribution (KGD) system which is built upon the agreement of the blockchain peers \cite{clKGC}. KGD are the blockchain peers that commence the process of device registration. It starts with system parameters and outcomes of the partial secret ($PS$). It eliminates the requirement of a trusted third party (TTP) such as the Certificate Authority of PKI. Before posting the transaction, IAVS RSU or sensors obtain public-private key pairs upon the completion of the registration process. The following part discusses how to source devices are registered to the blockchain network. Then how it verifies particular transactions submitted to it. 

\subsection{Registering RSU} At the beginning, multi-party IAVS stakeholders agree to build and share over the consortium blockchain (BC). Suppose, several RSUs cooperatively form the BC network that facilitates the key generating and  distribution (KGD) peers \cite{li2018blockchain}. Blockchain KGC peers broadcast the system parameters $(Y)$ all IAVS RSU have knowledge about. KGD peers keep their individual signer’s secret such as $S_1$,$S_2$...$S_n$ With the help of Edge computation capacity or its own ability, interested IAVS RSU creates their own secret value $X_1, X_2, X_3...X_j$ generates respective public keys using $X_j$ and the system parameter $Y$, where $j$ is the number of interesting devices at particular time $t$ and $n$ is the number of co-signing blockchain peers \cite{cons1}. 

\vspace{\baselineskip}

\noindent RSU server will contact the KGD with their identities $ID_1, ID_2,...ID_j$. Upon receiving the request, KGD will generate a partial private secret $PS_1,PS_2...PS_i$ for all requested devices and will cosign co-sign $ID_i$ and $PS_i$ using co-signers private key $S_n$. KGD sends the signed message back to the RSU Edge. The sensor device itself or edge node (e.g. Azure IoT edge or Dell Gateway) will verify if the message comes from the KGD, and if yes, it will generate private - private key pairs $(Pk_1, Pk_2,...,Pk_i , Sk_1,Sk_2,...,Sk_i)$  using $(PS_i,X_i,Y$. Note that only each PMU will be able to create the private key because it is the only entity who knows his private secrets $X_j$. \textbf{Alg}. \ref{Ch3_Alg2} illustrates the step by process with the necessary explanations.

\vspace{\baselineskip}

\begin{algorithm}

% functions

\SetKwFunction{setup}{setup}
\SetKwFunction{keyGen}{keyGen}
\SetKwFunction{genSk}{genSk}
\SetKwFunction{requestSend}{requestSend}
\SetKwFunction{genPS}{genPS}
\SetKwFunction{responseReceived}{responseReceived}
\SetKwFunction{multiSig}{multiSig}
\SetKwFunction{genPS}{genPS}
\SetKwFunction{verify}{verify}
\SetKwFunction{genPk}{genPk}

% input/ouput names
\SetKwInOut{Input}{Input}
\SetKwInOut{Output}{Output}

% caption
\caption{IAVS RSU registration with blockchain KGD.\label{Ch3_Alg2}}

\Input{%
		\xvbox{4mm}{$\xvar{ID}_j$} -- identities of the $j'th$ number of vehcile  \\
		\xvbox{2mm}{$\xvar{Y}$} -- system parameters \tcc*{prime numbers, primitive roots etc}
	  }
\Output{%
		\xvbox{12mm}{$\xvar{(Pk, Sk)}$} -- public and private key pairs \tcc*{for all devices at $t$}
		} 

  \BlankLine % blank line for spacing
  
  % start of the pseudocode
  \xvbox{12mm}{$\xvar{\setup}(1^\lambda)$} $\rightarrow (Y)$ \tcc*{system parameters $(Y)$ initialization}
  
  \For{$\xvar{ID}  \leftarrow$ $\xvar{ID}_j$ }{
  
  \xvbox{2mm}{$\xvar{porocedure \keyGen(Y, ID):}$}  \tcc*{key using system $Y$ and identities}

  \xvbox{2mm}{$\xvar{X}_{ j }$} $\;\leftarrow$ \genSk(Y,$ID_j$) \tcc*{IAVS generates own secret keys}
  
  \xvbox{2mm}{$\xvar{\requestSend(ID}_j)$}  \tcc*{IAVS send interests to join consoritum BC}
  
  \xvbox{4mm}{$\xvar{PS}_{ j }$} $\;\leftarrow$ \genPS($ID_j$) \tcc*{KGDs generates partial secret}
  
  \xvbox{12mm}{$\xvar{\multiSig}(S_n,ID_j,PS_j)$}  \tcc*{multi-sign using $S$ of $n$ cosigners}
    
  \xvbox{2mm}{$\xvar{\responseReceived(ID}_j)$}  \tcc*{IAVS receives $PS$ from KGD}
  
  \xvbox{14mm}{$\xvar{V\;\;[0,1,\xvar{$\bot$} ]}\leftarrow\;$$\xvar{\verify()}$}   \tcc*{verify the multisignatures}
    
  \If{$\xvar{V}\leftarrow$ $\xvar{1}$ }{

    \xvbox{4mm}{$\xvar{Sk}_j$} $\;\leftarrow$ \genSk(Y,$ID_j$,$X_j$) \tcc*{sets IAVS device private key}
    
    \xvbox{4mm}{$\xvar{Pk}_j$} $\;\leftarrow$ \genPk(Y,$X_j$) \tcc*{sets IAVS device public key}
  } % end for j	
  }

\end{algorithm}

\subsection{Transaction Verification and Preservation} Once RSUs are successfully registered to the KGD upon the certificate-less cryptography and multi-signature-based authentication, the RSU proceed further to send and store data. Usually, data gets transaction fashioned before sending it to the blockchain network. The transaction includes the identity of the RSU along with the action and timestamp at the time $(T)$ of action ($ACT$). There can be different types of actions such as $store$ data at a specific DHT address ( $ADS$), $update$ previously inserted data or $access$ permission of the particular data. To verify a transaction $T_x = (ID_j,T, ACT)$, the blockchain peers have to meet two conditions: \textit{i)} Either the public key ($PK_j$) obtained associates with the identity ($ID_j$), \textit{ii}) or any other public parameters can the signed transaction ($T_x$) be verified \cite{cl, multi}. 

\vspace{\baselineskip}
\noindent RSA (Rivest-Shamir-Adleman) based digital signature algorithm (DSA) or elliptic curve digital signature algorithm (ECDSA) can be used. Considering the lesser key-size facility, we opted for the ECDSA in our evaluation setup inside the apache Kafka framework of the hyperledger Iroha framework \cite{iot1}. 

\vspace{\baselineskip}
\begin{algorithm}

% functions
\SetKwFunction{create}{create}
\SetKwFunction{signTx}{signTx}
\SetKwFunction{castTx}{castTx}
\SetKwFunction{verID}{verID}
\SetKwFunction{verTx}{verTx}
\SetKwFunction{storeDHT}{storeDHT}

% input/ouput names
\SetKwInOut{Input}{Input}
\SetKwInOut{Output}{Output}

% caption
% \addto\captionsenglish{\renewcommand{\algorithmcfname{Alg.}}

\caption{IAVS Transaction (Tx) verification and storing.\label{Ch3_Alg3}}

\Input{%
		\xvbox{4mm}{$\xvar{T}_x$} -- IAVS transactions  \\
		\xvbox{2mm}{$\xvar{L}$} -- access control lists \\
		\xvbox{2mm}{$\xvar{}$$\sigma$} -- signaturues of the $T_x$ \\
		\xvbox{4mm}{$\xvar{ID}_j$} -- identities of the $j'th$ number of RSU Servers  \\
		\xvbox{2mm}{$\xvar{Y}$} -- system parameters \tcc*{prime numbers, primitive roots etc}
	  }
\Output{%
		\xvbox{17mm}{$\xvar{($V_{I_d}$,$V_{T_x}$, S)}$} -- set \& return verification and storing flag \textit{true}
		} 

  \BlankLine % blank line for spacing
 
  \xvbox{2mm}{$\xvar{\create \; :=  (ID,\;L,\;Tx,\;$\sigma$,\;ADS)}$}  \tcc*{creates $Tx$ using $L$ $ID$ and $ADS$}
  
  \xvbox{2mm}{$\xvar{\signTx(Tx},Sk)$}  \tcc*{sign creates transactions}
  
  \xvbox{2mm}{$\xvar{\castTx(Tx},\sigma)$}  \tcc*{broadcasts the original $Tx$ and the signed one}
  
  \For{$\xvar{Tx}  \leftarrow$ $\xvar{Tx}_i$ \tcc*{for all transaction $n\times T_x$} }{
  
  \xvbox{2mm}{$\xvar{$V_1 \leftarrow$ \verID(ID},Pk,Y)$}  \tcc*{verifies the identities ($ID$)}
  
  \xvbox{2mm}{$\xvar{$V_2 \leftarrow$ \verTx(Tx,ID},Pk,\sigma)$}  \tcc*{verifies the transactions ($Tx$)}

  \If{$\xvar{(V}_1\;||\;V_2)$ }{

    \xvbox{2mm}{$\xvar{S $\leftarrow$ \storeDHT(Tx, ID):}$}  \tcc*{store $Tx$ into DHT and set $S$ true}
  }
  }

\end{algorithm}

\vspace{\baselineskip}

\begin{figure}[ht]
\centering
\includegraphics[width= \linewidth]{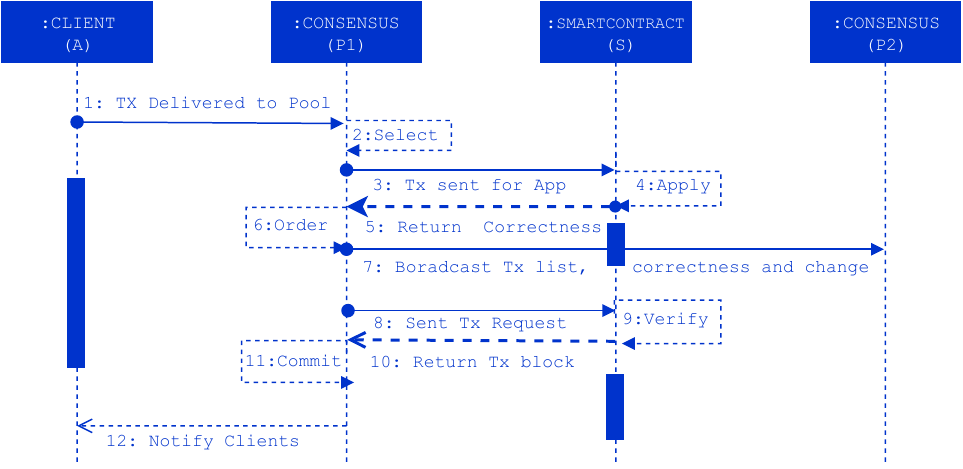}
\caption{Communication sequence of a typical IAVS within the blockchain network. It employs the cycle among client devices that submit data, consensus peers and smart contact}
\label{Ch3_Fig9}
\end{figure}

\noindent Here, the signature algorithm can be represented as a triple /4-tuple of probabilistic polynoimial-time algorithms ($G,S,V$) or ($G, K, E,D$) that includes generation ($G$), $signing(S)$, verification($V$), key-distribution($K$), encryption($E$) and decryption($D$) respectively. Upon successful verification, the address ($ADS$) is stored in the DHT while the pointer belongs to the blockchain peers who verify. The following \textbf{Alg}. \ref{Ch3_Alg3} shows how the mechanism happens. Besides, the identities $ID_j$, here the devices require the Access Control List (ACL) before Transaction ($Tx$) creation and signing ($\sigma$).The industry 4.0 devices along with the RSU Gateway are solely responsible to create the ACL list ($L$) in addition to signature ($\sigma$) generation and transaction($T_x$) publishing. However, the same $L$ will be required later to access data. The algorithm as shown in \textbf{Alg}. \ref{Ch3_Alg3}, outcomes three different flags ($V_1,V_2,S$) set after successful execution. If the identities belong to the derived public keys, $V_1:=true$, while the certificateless signature meets the condition as discussed earlier, ($V_2:=true$).The blockchain peers do the transaction ($T_x$) verification in response to the reception. Interchangeable verification procedure works in case of data access. Similarly, upon RSU data transactions ($Tx$) are written into the DHT, the third flag gets set,($S:=true$). After that, a new block is added to the blockchain and subsequently, the ledger gets updated including the $T_x$ Pointer ($Tp$). Figure \ref{Ch3_Fig9} shows the communication sequence among client, consensus and smartcontract.

\vspace{\baselineskip}

\subsection{Smart Contract and Consensus Implementation}
The implementation required writing chain codes (CC, special smart contract for Hyperledger blockchain) against the respective ledger. The initial chain codes provide authentication, access control and authorisation while others ensure logging besides the validation. Being a platform-independent platform Hyperledger supports any language to write its codes, however, because of relevant online resources, we preferred mostly $Go$ and in some test-cases $Java$. To adapt multi-signature-based certificateless environment after eliminating certificate authority (CA) and membership service provider (MSP), the dependencies of the open-source $shim$ package needed customization \cite{gdpr, cons1}. By default, it provides ledger/other CC accessing APIs, state variables or ($T_x$) context. Considering the data\_pointer represents the cipher-text of the IIoT data. Assuming an $encryption$ function $\mathbb{E}$ with public key ($Pk_j$):$ data\_pointer  = \mathbb{E}(mQA  Pk_j, device.id_j) $. A third-party entity with a shared private key ($Sk_j$) can decrypt the $device.id$ as well using  an opposite decryption function $\mathbb{D}$.$ device.id = \mathbb{D}(Sk_j,data\_pointer) $. The policy in the $IIoT-ledger1.1$ is simply defined as an access control list (ACL). Figure 9 depicts the communication among client devices (edge gateway), smart-contract and consensus protocols. Firstly, the transaction ($Tx$) is submitted to the blockchain network as a proposal using a smart contract. The SDK of the network provides the application environment to check if it is valid. Once valid it needs to be consented to by the consensus peers. In doing so it broadcast the $Tx$ among all collaborating peers of the consortium and updates the ledger \cite{hlf}.

\vspace{\baselineskip}
\begin{figure}[ht]
\centering
\includegraphics[width= \linewidth]{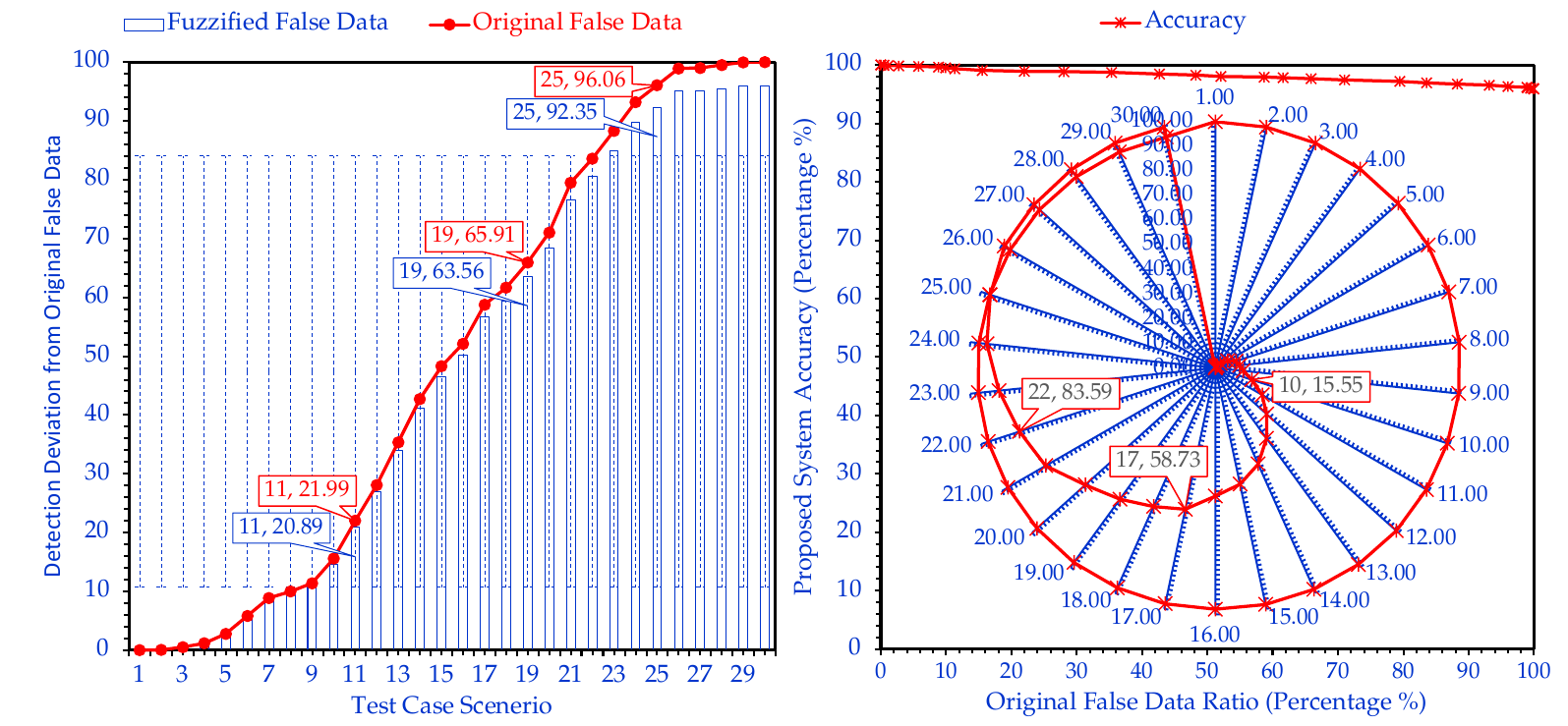}
\caption{Detection deviation and accuracy of the proposed AI-enabled detection technique. It shows corresponding data for selected 30 test cases }
\label{Ch3_Fig10}
\end{figure} 

\section{Evaluation and Result Analysis}
The proposed model was tested on \textit{FuzzyTech} simulation tool and the detected data and its preservation were purposefully verified on the consortium blockchain platform namely hyperledger fabric. The following section discusses the obtained result accordingly. To evaluate the proposed system accuracy we have implemented Mamdani fuzzy inference system (FIS) on a Windows 10 enterprise operating system working on \textit{Intel Core(TM) i5-7200U} CPU with 8 \textit{GB} RAM 2.50 and 2.71 \textit{GHz} capacity.

\subsection{Detection Accuracy}
\begin{figure}[ht]
\centering
\includegraphics[width= \linewidth]{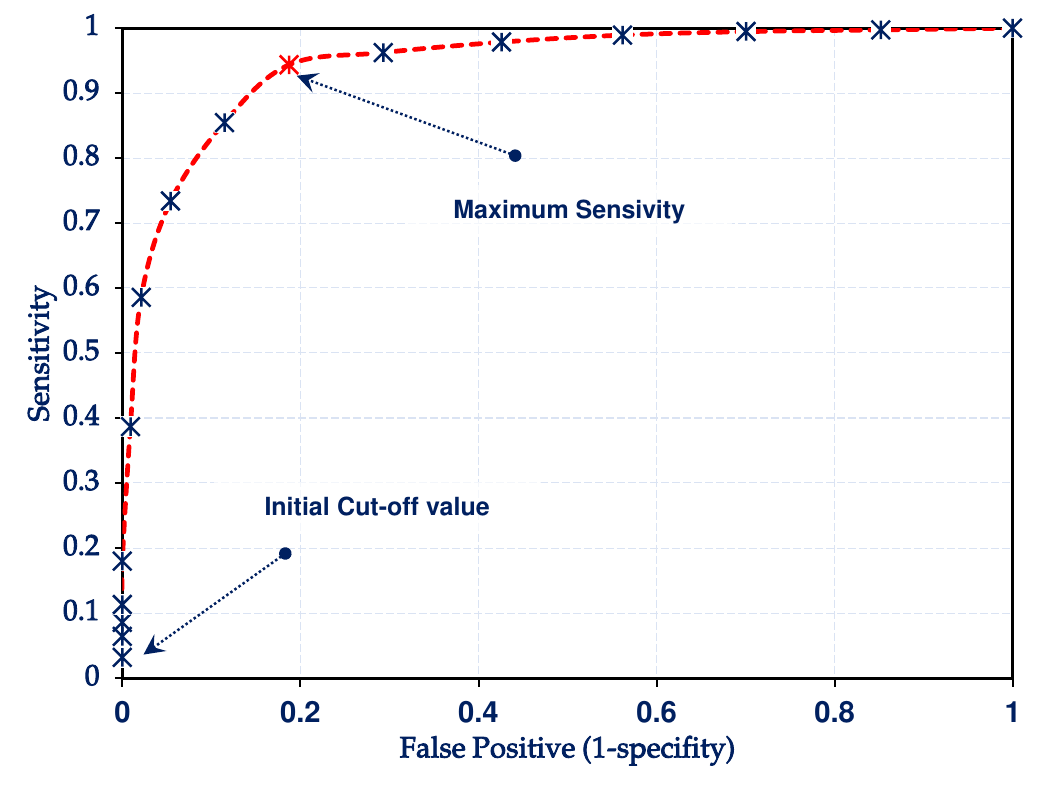}
\caption{Receiving operating characteristic (ROC) analysis of the proposed detection technique based on the considered test cases. It marks the initial cut-off and maximum sensitivity region of the proposed Fuzzy-enabled model.}
\label{Ch3_Fig11}
\end{figure}

The built system was debugged for several cases. Among all debugs, there were 30 test cases used to visualise the chart fuzzy system accuracy. Figure \ref{Ch3_Fig10} shows the detection trend of the system. The detection was made using the fuzzy input and respective MF based on the rules considered. The rule extraction section of the manuscript explains the notations used for each rule. The initial portion of Figure \ref{Ch3_Fig10} shows the detection deviation from the original injection of false data. The latter portion presents the accuracy of the system. Two rules were used based on the variation of Error and Weight; therefore, the accuracy shown is only the accuracy of the selected MFs, which actually differs for higher number rules. It shows that it has higher accuracy when IAVS RSUs have fewer anomalies, and the accuracy slightly decreases with a higher injection of false data. Further work will be undertaken to improve this finding. The corresponding receiver operating characteristic (ROC) curve (see Figure \ref{Ch3_Fig11}) presents the possible cut-offs for sensitivity and specificity among the $30$ test cases. It shows the system has maximum sensitivity with fewer false data, which portrays the most usable cut-off. The highest cut-off has the maximum true positive rate and the minimum false positive detection.

\vspace{\baselineskip}

\begin{table}[ht]
\caption{\label{Ch3_Tab2} The throughput (TP), success rate (SR) and delay latency (DL) for READ and WRITE operation of the IAVS transaction (Tx). The above values are calculated based on the workload (WL or Tx per second) as shown in the left-most column.}

\begin{tabular}{lllllll}
\toprule
    & \multicolumn{3}{c}{\textbf{READ}} & \multicolumn{3}{c}{\textbf{WRITE}} \\
    \midrule
   \textbf{WL} & \textbf{TP}     & \textbf{SR}     & \textbf{DEL}    & \textbf{TP}     & \textbf{SR}     & \textbf{DEL}     \\
100 & 100 & 9.7 & 0.01 & 100 & 9.7 &  0.01   \\ 
200 & 220 & 9.6 & 0.01 & 185 & 9.3 &  0.31   \\ 
300 & 340 & 9.5 & 0.01 & 175 & 7.8 &  1.48   \\ 
400 & 390 & 9.3 & 0.02 & 145 &  6.1      &   4.38      \\ 
500 &      470  &    8.8    &     0.05   &   85     &   4.9     &    5.25     \\ 
600 &     370   &    7.5    &     2.26   &   60     &   3.8     &    6.16     \\ 
700 &     330   &    6.5   &     5.43   &    40    &    3    &     7.43    \\ 
800 &     250   &    5.5   &     6.12   &    30    &    2    &     8.22    \\ 
900 &      170  &    4.5    &   6.41     &   20     &   1.1     &  8.94       \\ 
  1000  &     110   &    3.5    &    6.68    &   10     &   0.5     &  9.92      \\
   
    \bottomrule
\end{tabular}

\end{table}

\subsection{Blockchain Network Performance}
\begin{figure}[ht]
\centering
\includegraphics[width= \linewidth]{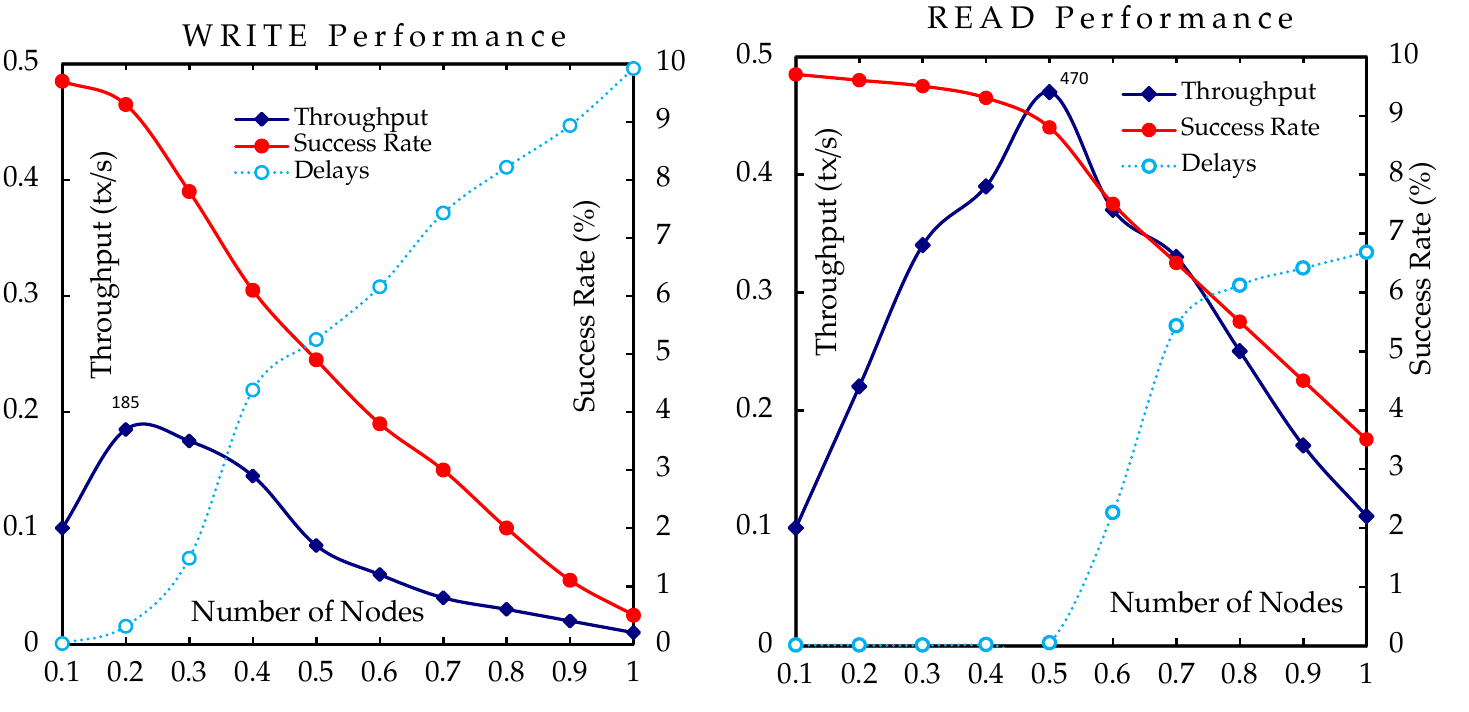}
\caption{READ and WRITE performance of the deployed blockchain network that securely record data and reputation and store data to the associated distributed hash table (DHT)}

\label{Ch3_Fig12}
\end{figure}

The HLF benchmarking results shows the performance based on four measurement metrics success rate ($\rho$), latency ($\Delta$ t and $L$) and the Throughput ($P$) and the resource consumption ($W$) for different test cases. Figure \ref{Ch3_Fig12} shows the system performance under a different number of workloads ($W$) ranging from 0.1k to 1k workload where the HLF network occupies two (02) chain codes, four (04) peer nodes and three (03) OSNs running on apache Kafka for practical byzantine fault tolerance (PBFT) consensus. As seen in the figure the $WRITE$ has 185 at 0.2k workload ($W$) with a maximum success rate of 93\% and an average delay of 5 seconds. On the other hand, $READ$ operation seems to have higher throughput (up to 470 at maximum) on a similar success rate at its best. The average delay seems to be half of the write's delay as the write has to incorporate OSNs on Apache Kafka. Table \ref{Ch3_Tab2} shows the throughput (TP), success rate (SR) and delay (DEL) latency of the blockchain deployment \cite{gdpr}.

\vspace{\baselineskip}
\noindent The benchmark evaluation explicitly illustrates that the setup configured has lower performance for a higher number workload ($W$) though the theoretical solution proves the consortium blockchain has significant adaptability for a higher number of nodes. As investigated deep inside, the local workload processing bottleneck affects throughput and latency. Hyperledger $T_x$ flow works demand enough responses against the submitted $T_x$ proposals, in case the responses are queued due to network overhead, bandwidth or processing loads consequences of the latency raising. On top of that, the general purpose workstation configuration slower the evaluation for higher workloads \cite{hlf}.

\section{Conclusion} % and Future Work}
\label{conclusion}
In today’s IAVS data integrity attacks like FDI are an ongoing concern. If the system has inaccurate data, any activities based on that anomalous data will be in vain and can result in operational failure, financial cost and loss of lives. The proposed blockchain and Fuzzy-enabled false data–detection system should help filter anomalous data before sending it for further processing. Communication between the RSU and storage devices happens with collaborative verification, which ensures the system’s security and data safety. The system obviates the PKI-driven trusted CA and the established centralised system. Thus, it can eliminate SPOF and single-party dependency. The respective evaluation and results show that the proposed model has comparatively higher accuracy. The blockchain network’s performance justifies the proposed model’s applicability for the RSU and Vehicles. Future scope includes improving the accuracy of the number of behaviour rules and justifying the scalability for massive networks.

\bibliographystyle{ACM-Reference-Format}
\bibliography{sample}

% \appendix









\end{document}